\RequirePackage{lineno}
\documentclass[aps,prl,twocolumn,floatfix,reprint]{revtex4-1}
\usepackage{graphicx}
\usepackage{color}
\usepackage{amssymb} 
\usepackage{amsmath} 
\usepackage{natbib}
\usepackage{soul}
\usepackage{titlesec}
\usepackage{physics}
\usepackage[utf8]{inputenc}
\usepackage{hyperref}
\hypersetup{
    colorlinks=true,
    linkcolor=blue,
    filecolor=magenta,      
    urlcolor=cyan,
    citecolor=blue,
}

\setcitestyle{super, numbers, square,}

\newcommand{\degree}{^{\circ}}

\usepackage{color}
\definecolor{pu}{rgb}{0,0.,0.55}  
\definecolor{rd}{rgb}{1,0,0} 
\definecolor{gr}{rgb}{0.6,0.6,0.6}

\begin{document}
\title{Valley population of donor states in highly strained silicon}

\author{B. Voisin$^{1,2}$}
\author{K.S.H. Ng$^{3,4}$}
\author{J. Salfi$^{3,5}$}
\author{M. Usman$^{6,7}$}
\author{J.C. Wong$^8$}
\author{A. Tankasala$^9$}
\author{B.C. Johnson$^{6,10}$}
\author{J.C. McCallum$^6$}
\author{L. Hutin$^{11}$}
\author{B. Bertrand$^{11}$}
\author{M. Vinet$^{11}$}
\author{N. Valanoor$^8$}
\author{M.Y. Simmons$^3$}
\author{R. Rahman$^3$}
\author{L.C.L. Hollenberg$^6$}
\author{S. Rogge$^3$}

\affiliation{{$^1$\, Silicon Quantum Computing, Sydney, NSW 2052, Australia}\\
{$^2$\, School of Physics, Sydney, The University of New South Wales, Sydney, NSW 2052, Australia}\\
{$^3$\, Centre for Quantum Computation and Communication Technology, School of Physics, The University of New South Wales, Sydney, NSW 2052, Australia}\\
{$^4$\, 5. Physikalisches Institut and Center for Integrated Quantum Science and Technology, Universit\"{a}t Stuttgart, 70569 Stuttgart, Germany}\\
{$^5$\, Department of Electrical and Computer Engineering, University of British Columbia, Vancouver, BC V6T 1Z4, Canada}\\
{$^6$\, Centre for Quantum Computation and Communication Technology, School of Physics, The University of Melbourne, Parkville, VIC 3010, Australia}\\
{$^7$\, School of Computing and Information Systems, Faculty of Engineering and Information Technology, The University of Melbourne, Parkville, 3010, Victoria}\\
{$^8$\, School of Materials Science and Engineering, The University of New South Wales, Sydney, NSW 2052, Australia}\\
{$^9$\,Electrical and Computer Engineering Department, Purdue University, West Lafayette, Indiana, USA}\\
{$^{10}$\,Quantum Photonics Laboratory, School of Engineering, RMIT University, Melbourne, Victoria 3000, Australia}\\
{$^{11}$\, Universit\'{e} Grenoble Alpes, CEA, LETI, 38000 Grenoble, France}}

\begin{abstract}
Strain is extensively used to controllably tailor the electronic properties of materials. In the context of indirect band-gap semiconductors such as silicon, strain lifts the valley degeneracy of the six conduction band minima, and by extension the valley states of electrons bound to phosphorus donors. Here, single phosphorus atoms are embedded in an engineered thin layer of silicon strained to 0.8\% and their wave function imaged using spatially resolved spectroscopy. A prevalence of the out-of-plane valleys is confirmed from the real-space images, and a combination of theoretical modelling tools is used to assess how this valley repopulation effect can yield isotropic exchange and tunnel interactions in the $xy$-plane relevant for atomically precise donor qubit devices. Finally, the residual presence of in-plane valleys is evidenced by a Fourier analysis of both experimental and theoretical images, and atomistic calculations highlight the importance of higher orbital excited states to obtain a precise relationship between valley population and strain. Controlling the valley degree of freedom in engineered strained epilayers provides a new competitive asset for the development of donor-based quantum technologies in silicon.
\end{abstract}

\maketitle

\section{Introduction}

\begin{figure*}[htp]
\begin{center}
\includegraphics[width=\textwidth]{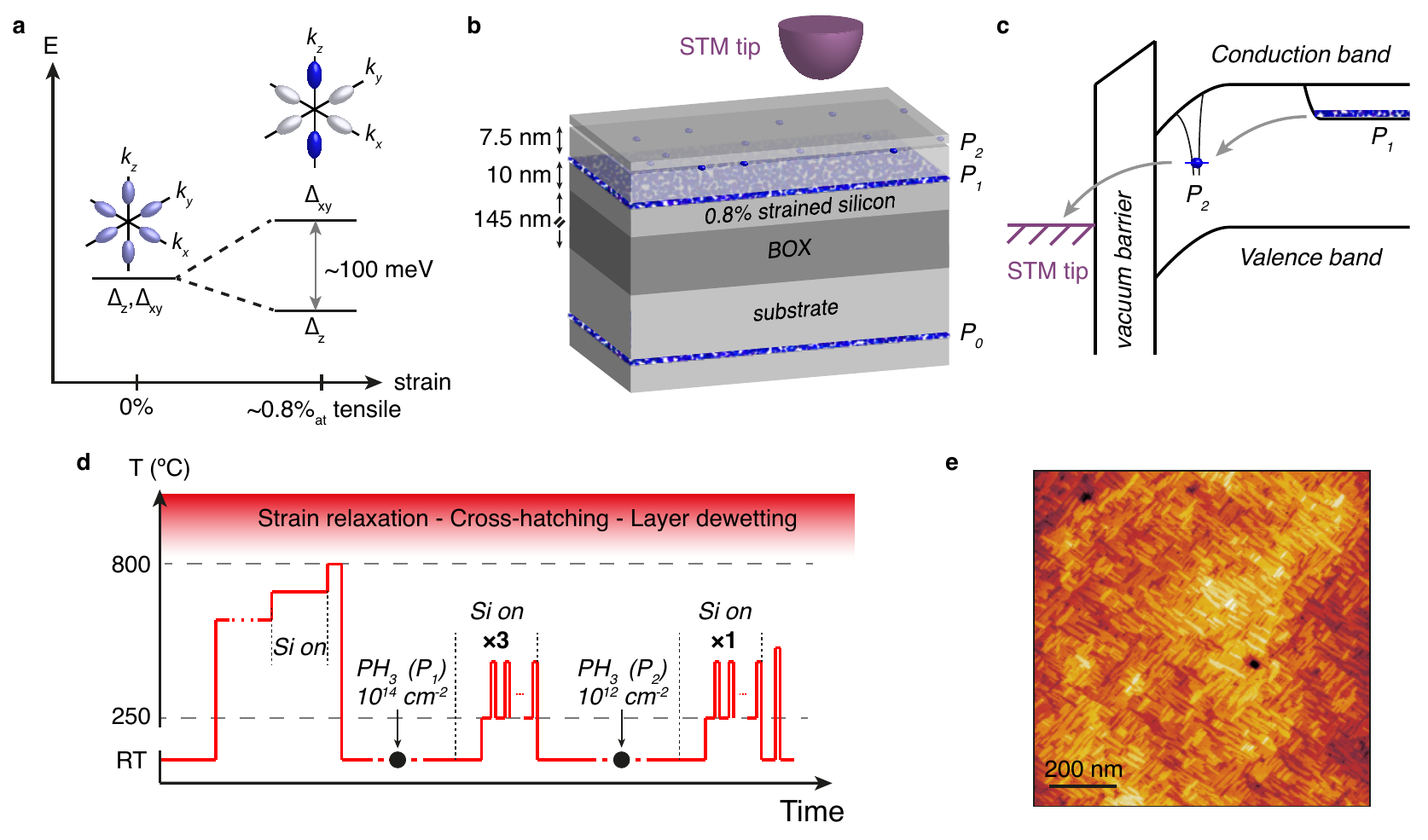}
\caption{\textbf{Sample doping and growth engineering. a,} The valley degeneracy of single donor ground states evolves from a six-fold to a two-fold degeneracy for large tensile strain, where the out-of-plane valleys $\Delta_z$ dominate with respect to the $xy$-valley states $\Delta_{xy}$. The schematics indicate a representation of the valley states in Fourier space. \textbf{b,} Schematics of the sample. Prior to in-situ fabrication and measurements, the sSOI substrate is doped with P atoms by ion implantation ($P_0$). Spatially resolved resonant tunneling is performed from a saturated (conductive) phosphorus layer which acts as a reservoir at low temperature ($P_1$), to an isolated P donor in the low dose layer  ($P_2$), to the STM tip. \textbf{c,} Diagram representing the electronic transport through a single donor. Transport occurs from the reservoir (first phosphorus doped layer, high dose), to a single donor found in the second phosphorus-dosed layer and then to the STM tip through the vacuum barrier. \textbf{d,} Low-temperature fabrication procedure of the sample. The thermal budget must not exceed 800$^\circ$C. Two phosphorus depositions, $P_1$ at high dose ($10^{14}$\,cm$^{-2}$ above the metal-insulator transition to create a reservoir) and $P_2$ at low dose ($10^{12}$\,cm$^{-2}$, to obtain isolated single donors), are separated by a 10\,nm silicon barrier with a growth procedure optimised to minimise segregation. A final 2.5\,nm growth follows $P_2$ to bury the isolated donors. \textbf{e,} Large scale topography after deposition of 10\,nm of silicon according to the procedure detailed in \textbf{d}.}
\label{fig1}
\end{center}
\end{figure*}

Strain is a modification of a crystal's lattice constant, which alters the electronic band structure and its degeneracies by breaking crystal symmetries. This effect, which can be captured by first principles methods~\cite{Richard2003,Munguia2008}, can modify the band energies, lift degeneracies, and reduce inter-band scattering~\cite{Niquet2012}. These properties have been used over the decades in microelectronics to enhance the mobility of nanoscaled MOSFETs~\cite{Yu2008,Niquet2012} as well as the luminescence of solar cells or 2D material systems~\cite{Wu2016,Deng2018}. By extension, strain also alters the quantum states of electrons bound to single dopants, defects or quantum dots~\cite{Hanson2007} which can be isolated and controlled in nanoscale electronic devices, since these states are built upon band states of the host material. Strain is invariably present to some degree in nanoelectronic devices for classical and quantum technologies, notably because of the material stack, of the presence of interfaces and nearby doping. Therefore, it is essential to develop tools that are able to probe the impact of strain at the nanoscale in order to understand and mitigate, but also possibly control and leverage strain in new architecture designs~\cite{Zhang2006,Thorbeck2015,Salfi2016}. In particular, strain lifts the valley and heavy/light hole degeneracies of silicon~\cite{Koiller2002,Zwanenburg2013} (see Fig.~\ref{fig1}a), with direct implications for exchange and hyperfine interactions~\cite{Koiller2002,He2019,Usman2015,Franke2015} as well as dipole and possible quadrupolar moments~\cite{Salfi2016,Asaad2020} of single dopants, all core properties that enable quantum information processing.\\

The sensitivity of electron spin resonance spectroscopy has been used to detect shifts in the hyperfine frequencies induced by strain on dopants in silicon~\cite{Wilson1961,Huebl2006,Lo2015,Mansir2018}. However, this technique averages the impact of strain over microscopic areas and over a large number of dopants, while the local impact of strain in a nanostructure and on a single quantum state is of interest for quantum applications based on individual dopants. Instead, microscopy techniques are well suited to study strain effects at the nanoscale. Whilst scanning transmission electron microscopy~\cite{Cooper2012} and atom probe tomography are able to detect single atoms in nanoscale devices~\cite{Koelling2017,Ishikawa2020}, scanning tunneling microscopy (STM) presents the advantage of giving direct access to wave function information~\cite{Yakunin2007}. This technique was recently used to probe the valley population of single donors in natural silicon~\cite{Salfi2014,Usman2016}. Here, we extend this work to spatially resolve the wave function of an electron bound to a single phosphorus atom embedded in a thin layer of highly strained silicon compatible with atomic precision lithography~\cite{Fuechsle2012}.\\

In this work, we report that the complex oscillating pattern which was observed in these images in the unstrained case due to the presence of all six valleys~\cite{Salfi2014,Usman2016} vanishes because of the valley repopulation effect. Using a combination of theoretical models for the exchange interaction, we show how this dominant $z$-valley donor wave function can be used in the context of atomically precise qubit devices to obtain robust exchange and tunnel couplings for any in-plane orientation and down to 5\,nm separation. A Fourier transform analysis highlights the remaining presence of an $xy$-valley population, which is unexpected according to effective mass theories for such large strain~\cite{Koiller2002}. The presence of this remaining in-plane valley population observed experimentally is in agreement with atomistic tight-binding (TB) simulations~\cite{Usman2015a}, which notably take into account high orbital excited states~\cite{Lansbergen2011,Tankasala2018} and have been benchmarked against ab initio density functional theory~\cite{Usman2015}. Our work provides a quantitative understanding of the effect of strain on valley population, relevant for the development of robust two-donor coupling in silicon.\\
 
\section{Sample fabrication and structural analysis}

\begin{figure}
\includegraphics[width=8.5cm]{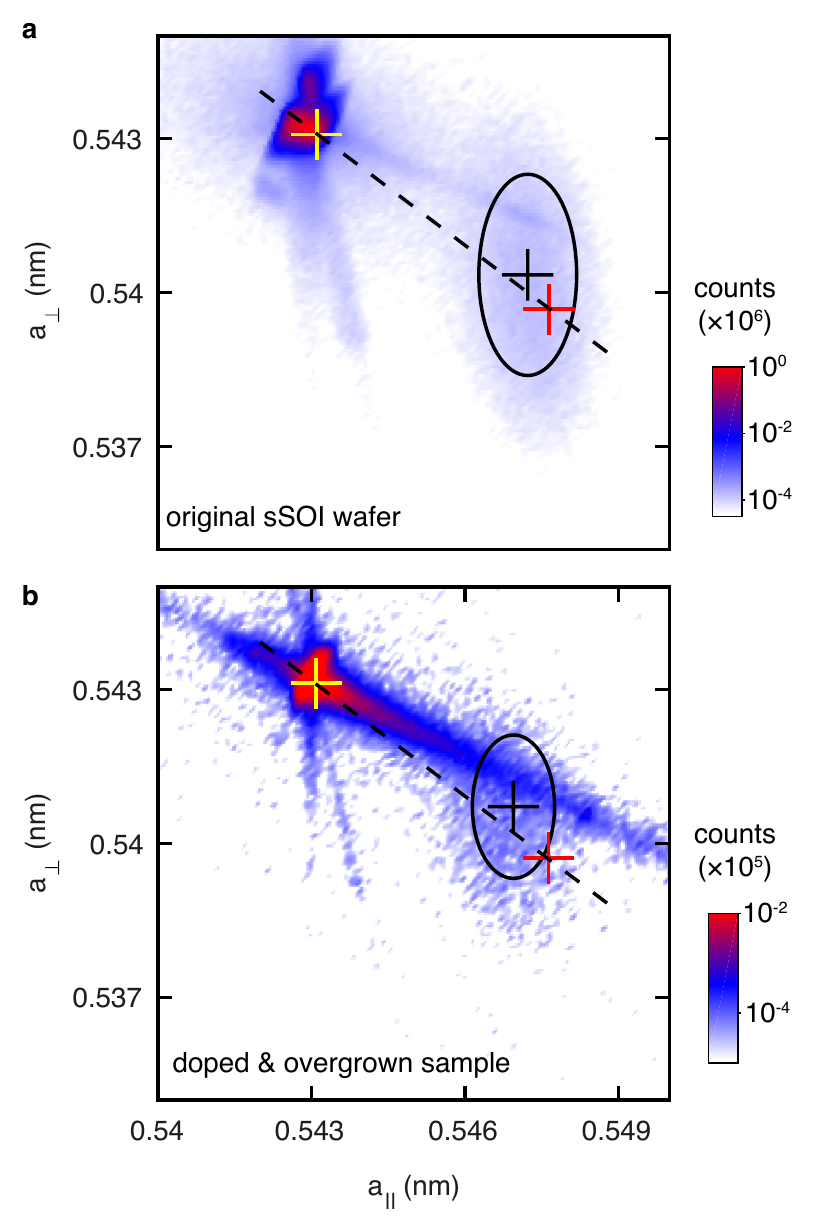}
\caption{\textbf{X-ray diffraction of thin strained silicon epilayers. a,} 2D ${2}\theta{-}\omega$ map taken along [224], reconstructed in real space, of the bare wafer. The black dashed line corresponds to the expected relationship between $a_{\parallel}$ and $a_{\perp}$ from elasticity theory, see equation~\ref{elast1}. The yellow cross indicates the position of the maximum, which correpsonds to the unstrained substrate and is used as a reference of the relaxed lattice constants. The strained layer appears as a 2D ellipse centred (black cross) very close to the expectation value for a layer grown on a $\rm{Si_{0.8}Ge_{0.2}}$ wafer (red cross). The black ellipse denotes the variance of a Gaussian fit of the strain ellipse. The other diagonal features are spurious interference from the measurement setup. \textbf{b} Similar X-ray diffraction measurements and analysis were performed on the sample that was spectroscopically measured at low temperature after growth and doping. The strain has remained after overgrowth and phosphorus incorporation with very limited relaxation.}
\label{fig2}
\end{figure}

The sample fabrication procedure achieves three main objectives. The native oxide needs to be removed without dewetting the epilayer nor relaxing the strain, a flat surface must be obtained to enable STM spectroscopy, and dopants must be embedded in the originally undoped strained epilayer to induce a single electron tunneling framework on isolated ones. The fabrication of our sample starts from a commercial 300\,mm wafer from Soitec with a 10\,nm-thick layer of highly strained silicon~\cite{Hartmann2003,Hartmann2007}. This silicon layer was grown on a $\rm{Si_{0.8}Ge_{0.2}}$ substrate to induce the desired amount of strain, and subsequently transferred on a silicon oxide substrate according to the Smart Cut process\texttrademark. Importantly, the strain should be preserved throughout the fabrication procedure which is achieved by implementing a low-temperature fabrication budget.\\

The substrate was first implanted at 3.5\,$\rm{\mu}m$-depth with phosphorus atoms at an energy of 6\,MeV and a dose of $10^{15}$\,cm$^{-2}$ in order to facilitate the annealing steps required during the {\it in-situ} fabrication procedure. Normally for such devices, this {\it in-situ} preparation starts with a 10\,h-anneal at 600$\rm{\degree{C}}$ followed by a high-temperature flash annealing (typically above 1050$^\circ$C) to remove the native oxide~\cite{Salfi2014}. However, this high temperature anneal cannot be performed here with a thin strained layer because it results in cross-hatching due to the strain variation across the wafer, which is notably unsuitable for STM lithography~\cite{Lee2014}. This high-temperature flash also incurs a high risk of dewetting the thin strained layer. Instead, we adopted a low-temperature recipe~\cite{Zhang2006} where the sample is kept at 700$\rm{\degree{C}}$ facing a silicon deposition source calibrated to a rate of 0.5\,$\rm{ML.min^{-1}}$ for one hour. This was followed with a 2\,min-anneal at 800$\rm{\degree{C}}$ to ensure a high-quality $2{\times}1$ surface reconstruction.\\

A schematic of the final device is shown in Fig.~\ref{fig1}b, with the corresponding electronic transport diagram shown in Fig.~\ref{fig1}c. Similar to previous work~\cite{Salfi2014,Voisin2015,Usman2016,Salfi2018}, a tunnel current occurs from the highly doped reservoir labelled $P_1$ to an isolated donor found in the lightly doped layer (labelled $P_2$) and then to the STM tip. The temperature of the fabrication process, starting from the initial step where the native oxide is removed, is described in Fig.~\ref{fig1}d. A saturation dose of phosphorus dopants $P_1$, which will be conductive and act as a reservoir at low temperature, is deposited and incorporated at the surface, followed by a silicon growth of 7.5\,nm. The growth is divided in three equivalent sub-growths of 2.5\,nm, where the first nanometer is grown at room temperature and the remaining 1.5\,nm grown between 250$^\circ$C and 400$^\circ$C, which is designed to minimise segregation whilst preserving a flat surface for STM studies~\cite{Keizer2015,Koch2019}. A second phosphorus deposition $P_2$ follows, with a lower dose of $10^{12}$\,cm$^{-2}$, designed to be able to randomly find isolated donors. A second silicon growth of 2.5\,nm buries the isolated donors, followed by a 10\,s-anneal at 600$^\circ$C to further flatten the surface, which is finally hydrogen passivated. A large scale topography image is shown in Fig.~\ref{fig1}e, which shows an excellent surface quality, with low defects with large terraces (larger than 20\,nm) suitable for STM spectroscopy and amenable to STM lithography to fabricate atomically precise qubit devices~\cite{Fuechsle2012, Kranz2020}.\\

\begin{figure*}[htp]
\begin{center}
\includegraphics[width=\textwidth]{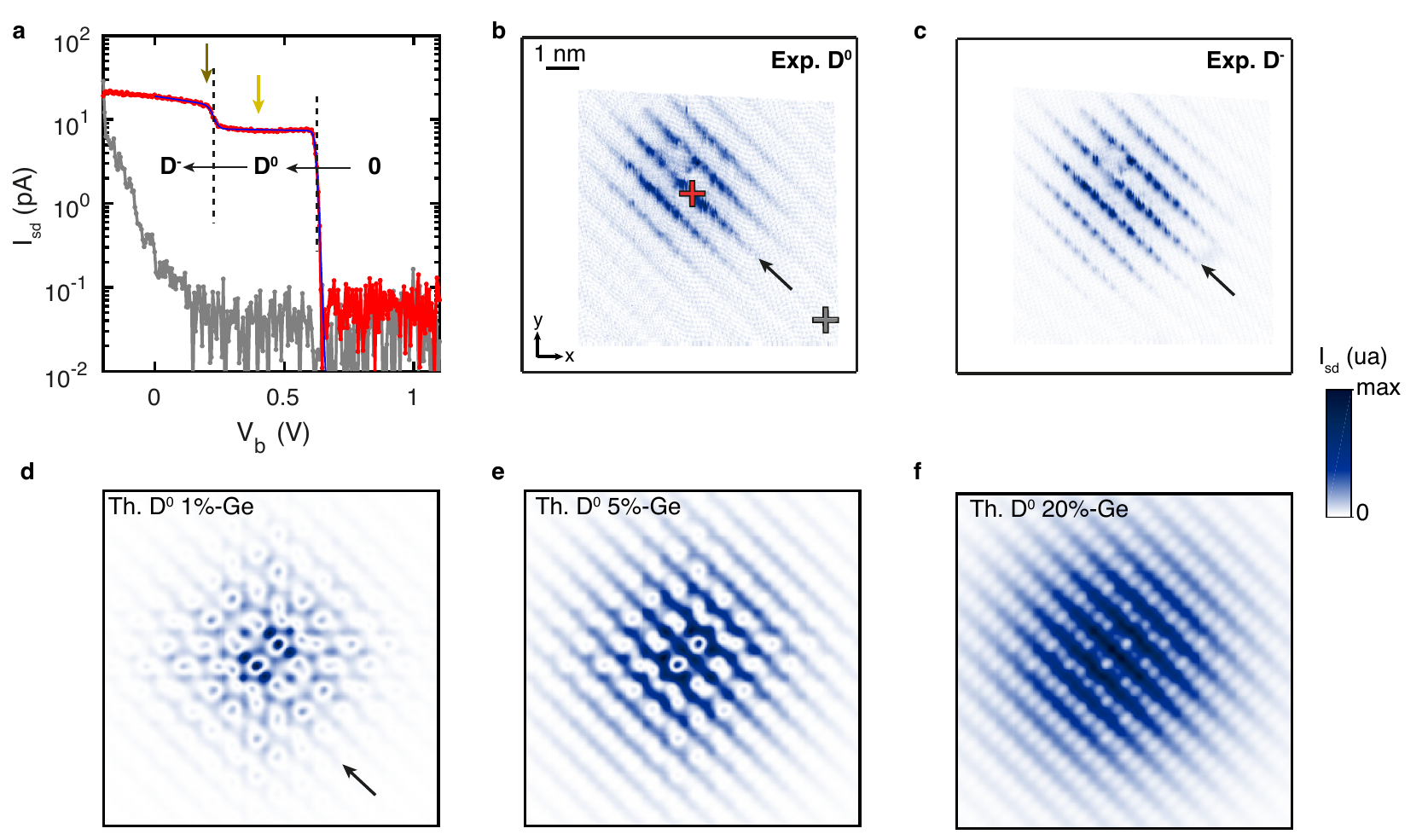}
\caption{\textbf{Single donor spectroscopy and wave function imaging. a,} Spectroscopy performed at 4\,K, away (grey) and over a single donor (red) embedded in the strained epilayer. Over the donor, two resonances appear below the onset of direct transport from the valence band states of silicon to the tip, which correspond to the addition of two electrons on the donor. The blue line is a fit to the current assuming a thermally broadened regime, from which the charging energy can be extracted. \textbf{b,} STM image of the current recorded at $V_b{=}0.4$\,V, corresponding to imaging the first electron transition ($D^0$ state). The red and grey crosses indicate where the grey and red spectra shown in {\it a} were taken. The scale and color bars are common to all the plots. \textbf{c,} STM image of the current recorded at $V_b{=}0.2$\,V, corresponding to imaging the second electron transition ($D^-$ state). Both the $D^0$ and $D^-$ images only show maxima along the dimer rows of the $2{\times}1$ reconstructed surface (indicated by the black arrows), which indicate the prevalence of the $z$-valleys in the donor ground state. \textbf{d-f,} Theoretical STM images using the states of a 4.5$a_0$-deep donor obtained from atomistic simulations in strained silicon originating from 1\%, 5\% and 20\%-Ge content in the buffer substrate, respectively. The pattern observed in the 1\% theoretical image, where  maxima are distributed along and across the dimers, relates to the presence of the $xy$-valleys. As the strain increases, this pattern evolves to only leave maxima aligned along the dimer rows. }
\label{fig3}
\end{center}
\end{figure*}

To confirm that we successfully inhibited strain relaxation by limiting the thermal and growth budget, we performed X-ray diffraction measurements through the overgrown layer. A ${2}\theta{-}\omega$ X-ray 2D map was taken along the [224] crystallographic axis for the bare sSOI wafer first, which is reconstructed in real space in Fig.~\ref{fig2}a. The lattice coordinates were corrected in order for the map maximum (yellow cross) to match the lattice constant of the sSOI strain-relaxed handling wafer in all three dimensions. The amount of germanium in the virtual substrate Si$_{1-x}$Ge$_x$ determines the resulting amount of strain in the silicon epilayer, which can be predicted from linear elasticity theory using the lattice constant mismatch between the two materials ($a_{Ge}{=}0.5658$\,nm and $a_{Si}{=}0.5431$\,nm). The resulting in-plane $a_{\parallel}$ and out-of-plane $a_{\perp}$ silicon lattice constants with respect to the unstrained case are:

\begin{equation}
\begin{gathered}
a_{\parallel}=(1-x)a_{Si}+xa_{Ge} \\
a_{\perp}-a_{Si}=2(c_{12}/c_{11})(a_{Si}-a_{\parallel})\sim0.75(a_{Si}-a_{\parallel})
\label{elast1}
\end{gathered}
\end{equation}

\noindent where $c_{11}$ and $c_{12}$ are the elastic constants of relaxed silicon~\cite{Windl1998}. Since $a_{Ge}{>}a_{Si}$, growing a silicon epilayer on Si$_{1-x}$Ge$_x$ results in a tensile in-plane strain, i.e. $a_{\parallel}{>}a_{Si}$ and $a_{\perp}{<}a_{Si}$,. We define $\epsilon_{\parallel}$ the in-plane strain (i.e., the relative change in the in-plane lattice constant), with $\epsilon_{\parallel}{=}0.8$\% for $x{=}0.2$. The black dashed line represents the linear relationship between the in-plane $a_{\parallel}$ and out-of-plane $a_\perp$ lattice constants following Eq.~\ref{elast1}. The thin strained silicon epilayer is evidenced as an ellipsoidal signal, which is centred very close to the expected location for a silicon layer grown on a 20\%-rich germanium substrate with $a_\parallel{=}0.5476$\,nm and $a_\perp{=}0.5397$\,nm (red cross). A 2D Gaussian fit of this signal (black ellipse, see Supporting Information section S1) yields a strain value $\epsilon_\parallel^{bare}{=}0.76{\pm}0.18$\,\% (black cross). A similar measurement and analysis was performed on the overgrown and dosed sample, as shown in Fig~\ref{fig2}b. A similar signal can be observed for this sample close to the expected strain value, and the same fitting analysis yields $\epsilon_\parallel^{exp}{=}0.71{\pm}0.15$\,\%. These results demonstrate that the strain is maintained during our anneal, growth and dosing procedure.\\

\begin{figure}[htp]
\begin{center}
\includegraphics[width=8.5cm]{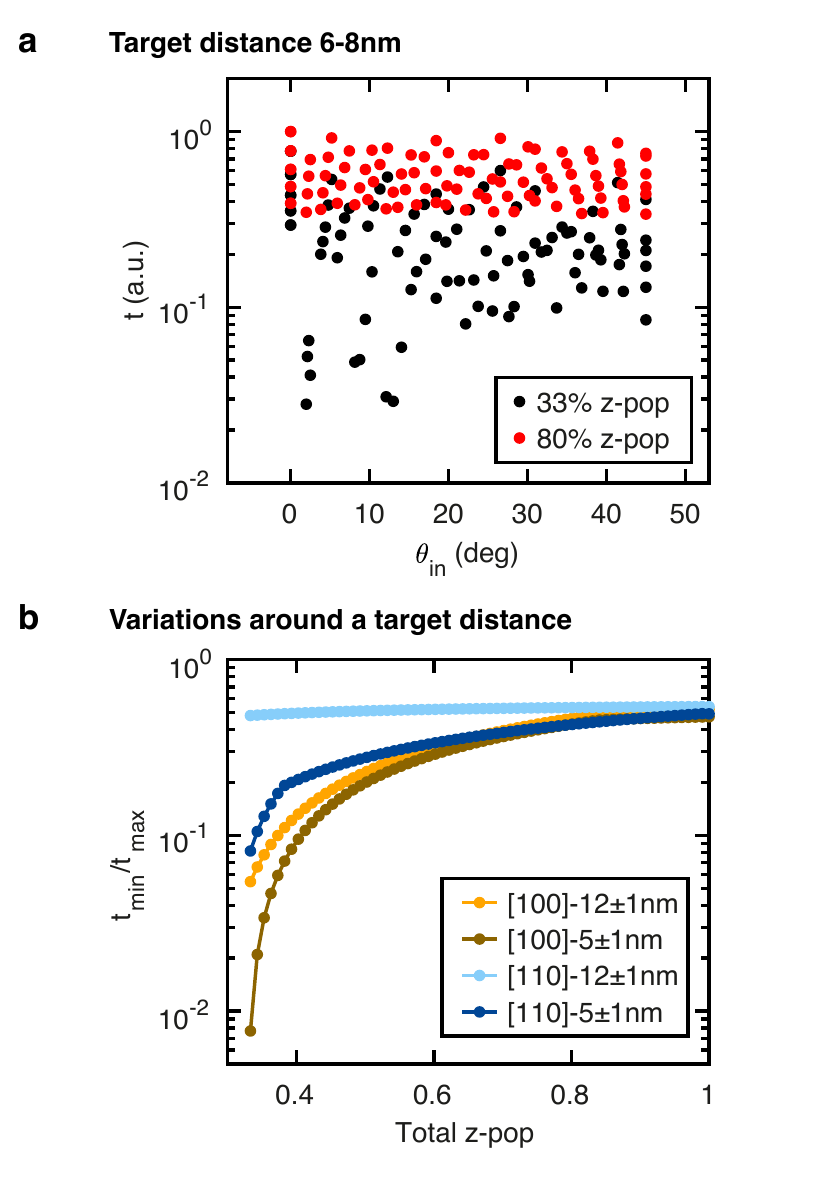}
\caption{\textbf{Valley population and tunnel coupling robustness to dopant position. a,} Normalised tunnel coupling values calculated for all the in-plane inter-donor positions between 6 and 8\,nm, for a $z$-valley population of 33\% (black points) and 80\% (red points). The angle $\theta_{in}$ is defined from the [100] crystallographic direction. The tunnel coupling becomes isotropic upon strain with a reduction of the large variations close to the [100] crystallographic axis. \textbf{b,} Ratios of the minimum to maximum tunnel coupling values found within a $\pm{1}\,$nm in-plane neighbourhood as a function of the $z$-valley population, for two target distances (5 and 12\,nm) and two target directions ([100] and [110]). Valley interference-induced variations are quenched irrespective of the distance and orientation for a $z$-valley population beyond 70\%.}
\label{fig5}
\end{center}
\end{figure}

\section{Spectroscopy and wave function imaging of a strained donor}

The objectives of this section are to spectroscopically evidence the presence of single donors addressable in the single electron tunneling regime, before spatially resolving their charge distribution and analysing these images in the context of the strain applied to the crystal. Donor states can be evidenced spectroscopically at 4.2\,K as they yield resonances below the direct transport onset from silicon valence band states to the STM tip~\cite{Salfi2014,Usman2016,Voisin2020}. This onset can be measured away from a single donor as shown in Fig~\ref{fig3}a-b. Here, it occurs at $V_b{\sim}0.2\,V$, which is offset compared to previous results~\cite{Salfi2014,Usman2016,Voisin2020} in unstrained silicon where it is found around $V_b{\sim}-0.8$\,V. This offset is attributed to an electrical short between the silicon epilayer and the handling wafer on the sample holder. Two extra resonances, at $V_b{\sim}0.6$\,V and $V_b{\sim}0.2$\,V respectively, can be observed when tunneling over a single dopant occurs (red line), which correspond to adding respectively the first (called $D^0$ state) and the second electron (called $D^-$ state) on the donor, the latter being close to the onset of direct transport from the valence band states~\cite{Salfi2014,Voisin2015}. This double resonance can be fitted to a thermal broadening model from which lever arms of $0.065{\pm}0.009$ and $0.072{\pm}0.003$ are obtained for the $D^0$ and $D^-$ resonances, respectively, as well as a charging energy of $14{\pm}2$\,meV (See Supporting Information section S2) between the two charge states. This value is low compared to the charging energy of a single phosphorus donor in bulk silicon (about 47\,meV~\cite{Fuechsle2012}). This difference can be explained by the presence of a metallic reservoir 10\,nm away~\cite{Salfi2018} and the strain environment which leads to a slight increase in the wave function envelope extent. We have performed full configuration interaction (FCI) calculations of the $D^-$ state following an atomistic tight-binding simulation of $D^0$. These calculations take into account both the presence of the strain in the lattice and of a screening metallic reservoir~\cite{Tankasala2018}, and find that the charging energy of a 4.5\,$a_0$-deep donor is decreased to 23\,meV for 20\%-Ge content strain at zero externally applied electric field, in reasonable agreement with the experimental value mentioned above.\\

\begin{figure*}[htp]
\begin{center}
\includegraphics[width=\textwidth]{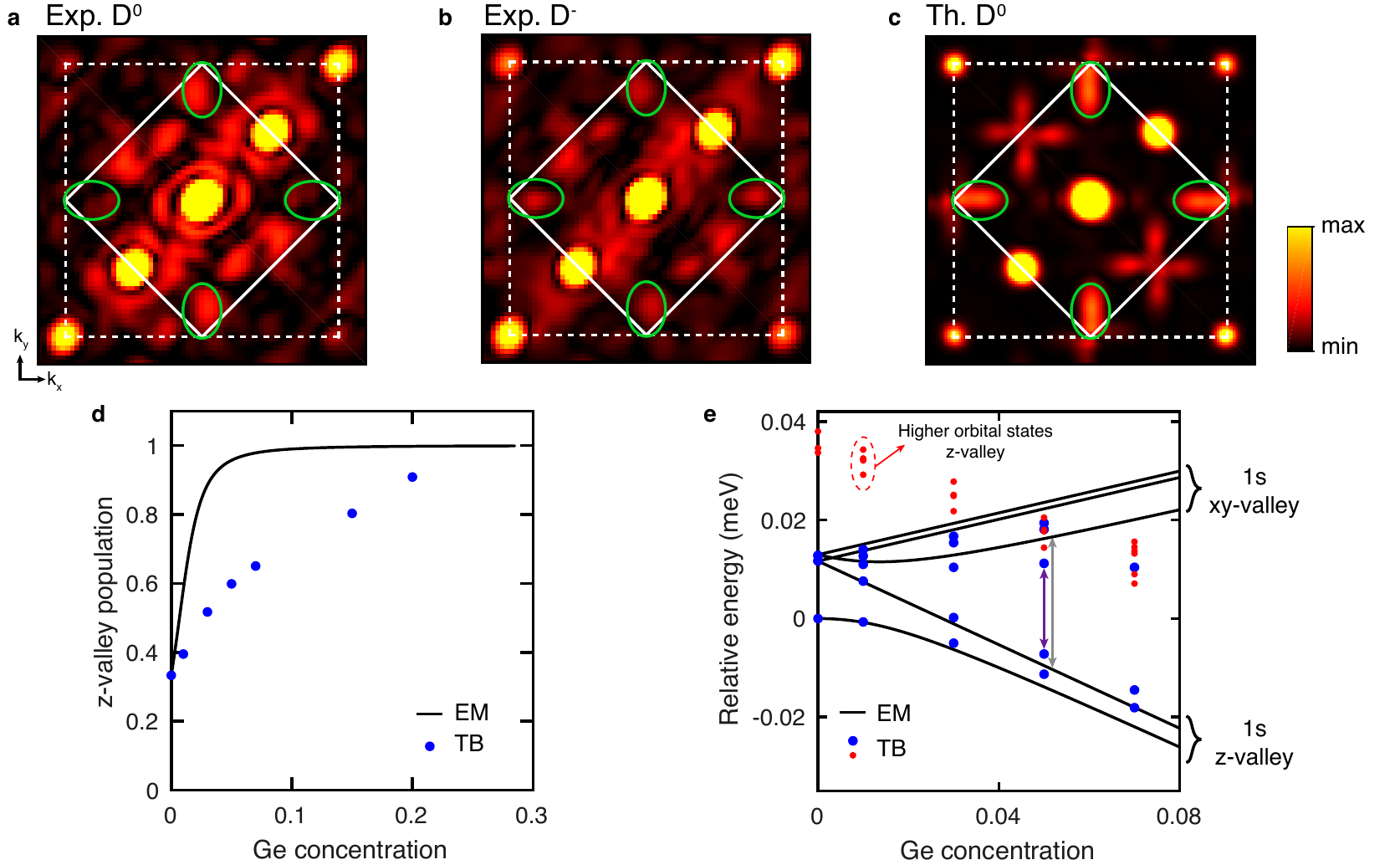}
\caption{\textbf{Fourier analysis and residual $xy$-valley population. a-b,} Fourier transform of the $D^0$ and $D^-$ images, respectively, centred on the first Brillouin zone (solid white lines). The signal circled within the green ellipses at $k_{\mu}{\sim}0.81k_0$ indicate a reminiscent $xy$-population of the donor ground state. \textbf{c,} Fourier transform of the theoretical STM image of the $D^0$ state for 20\%-Ge, which also shows a clear signal at $k_{\mu}{\sim}0.81k_0$. \textbf{d,} Evolution of the $z$-valley population as a function of strain expressed in Ge content, for an EM model (black line) and a full TB calculations (blue dots). The EM model yields a sharper prevalence of the $z$-valley population compared to TB. This indicates that the influence of higher excited states is required to maintain a finite $xy$-valley population for large strain as observed experimentally. \textbf{e,} State energies as a function of strain, for the EM model (black lines, $1s$-manifold) and TB calculations (blue dots for the states matching the $1s$-manifold of EM, red dots for higher orbital states). Within the $1s$ manifold, the $z$-valley states become favourable for both EM and TB, but  the energy splitting between the $z$-valley and the $xy$-valley states (grey arrow for EM, purple arrow for TB) is reduced in the case of TB, which slows down valley repopulation.}
\label{fig4}
\end{center}
\end{figure*}

A double-pass technique~\cite{Salfi2014} is used to image the charge distribution of the single donor evidenced in spectroscopy. The tip is first stabilised in topography, before setting the bias in the gap in a second pass where only the donor density of states contribute to the tunnel current. Following a drift correction protocol performed on the lattice topography image of the first pass, the charge distribution images are shown in Fig.~\ref{fig3}b and c, respectively for the $D^0$ state taken at $V_b{=}0.4$\,V and for the $D^-$ state taken at $V_b{=}0.2$\,V. In the unstrained case, the donor image reflects interference between waves at the valley and lattice frequencies, and the pattern strongly depends on the lattice plane at which the donor sits~\cite{Kohn1955,Salfi2014,Saraiva2016,Usman2016}. In contrast, the strained case studied here shows that the maxima of the charge distribution are aligned along the dimer rows. Following our previous work~\cite{Usman2016}, we have computed theoretical STM images based on a multi-million atom TB simulation of the donor $D^0$ state at a depth of 4.5\,$a_0$ for several amounts of strain. At 1\% (in Ge content), as shown in Fig.~\ref{fig3}d, the image reveals a complex symmetry with a set of oscillating patterns whose maxima are distributed both along and across the dimer rows. This is characteristic of the presence of all six valleys and of lattice frequencies. Going to 5\% and 20\%, as shown respectively in Fig.~\ref{fig3}e-f, this oscillating pattern progressively vanishes leaving a signal aligned with the surface dimer rows in agreement with the experimental images. This behaviour is consistent for any dopant depth (See Supporting Information section S3) to give confidence that the single donor experimentally imaged was embedded in a silicon layer strained to above 15\% equivalent Ge-content.\\

\section{Valley population impact on tunnel and exchange couplings.}

The change in the pattern observed in the STM images links to a change in the valley population. This effect has important consequences for dopant electronic properties, which we investigate theoretically in this section. We focus here on interactions based on the direct overlap between donor wave functions, such as tunnel and exchange couplings. These quantities can be very sensitive to the exact position of the dopants in the lattice because of the presence of interference at the valley frequency, an effect modulated by the anisotropy of the donor envelope that originates from the mass anisotropy in the conduction band of silicon~\cite{Koiller2001,Voisin2020}. Below, we investigate the effect of the valley population on these interactions in the context of atomically precise donor qubit devices, where donors are placed within the same crystallographic plane~\cite{Fuechsle2012,He2019}. We start from the Heitler-London expression of the exchange coupling using an effective mass description of the single electron wave functions, and of the relationship between tunnel and exchange couplings~\cite{Koiller2001,Voisin2020}:

\begin{equation}
\begin{gathered}
J(\vec{R})=\sum_{\substack{\mu,\nu= \\ \pm \{x, y, z\}}} \alpha_\mu^2\alpha_\nu^2j_{\mu\nu}(\vec{R})\cos(\Delta\phi_{\mu\nu}(\vec{R}))\\
t{=}\sqrt{JU}/2
\label{exc}
\end{gathered}
\end{equation}

\noindent where $\alpha_\mu$ is the amplitude of the valley population, $j_{\mu\nu}$ the envelope weight or overlap between the orbital part of valleys $\mu$ and $\nu$, and $\Delta\phi_{\mu\nu}$ a geometric phase difference term between the two valleys which depends on the distance between the two atoms and the valley momentum, and $U$ the on-site Coulomb repulsion energy of the $2e$-state which we consider to be constant. This exchange expression is in good agreement with full configuration exchange calculations down to inter-donor distances of 5\,nm~\cite{Tankasala2018,Voisin2020} (see Supporting Information section S4). In Eq.~\ref{exc}, the valley population gives an extra weight to each interfering envelope term $j_{\mu\nu}$. Their impact is shown in Fig.~\ref{fig5}a, where all the tunnel coupling values corresponding to in-plane donor positions between 6 and 8\,nm are plotted for two different valley populations. In the unstrained case, i.e. a $z$-valley population of 33\% ($\alpha_z{=}\alpha_x{=}\alpha_y{=}1/\sqrt{6}$), clear variations of over an order of magnitude close to the [100] axis can be observed, which are due to the sensitivity to $y$-valley interference, while the [110] axis is more resilient~\cite{Voisin2020}. Going to a $z$-valley population of 80\%, these large variations vanish to make the tunnel coupling isotropic in the plane, and we also note a relative increase of the maximum values. Both effects arise since the wave function is mainly concentrated in the $z$-valley and the $j_{zz}$ term is fully constructive, where less amplitude of the wave function is lost through destructively interfering terms in building the overlap. Making the direct interactions isotropic in the plane opens a way for quantum simulation architectures based on donor arrays with non-square geometries, where the donors are not only positioned along the [110] or $[1\bar{1}0]$ axis~\cite{Georgescu2014,Salfi2016,Le2020}.\\

We further investigated the impact of strain on the exchange variations caused by small changes in atom position. For two target orientations ([100] and [110]) and two target distances (5 and 12\,nm), we calculated the tunnel couplings values corresponding to all possible inter-donor positions within a 1\,nm in-plane neighbourhood. The ratios of the minimum to the maximum value are plotted in Fig.~\ref{fig5}b as a function of the total $z$-valley population. Along [110] the variations remain below a factor of 10 for small target distance and change in valley population. However, large variations are predicted along the [100] axis due to the inherent presence of destructive positions close by, for any target distance. Strain provides a way to overcome these variations. Overall, the impact of valley interference on tunnel coupling, and thus exchange coupling as well, is quenched for a $z$-valley population beyond 70\% leaving variations only due to changes in the donor envelope overlap. A remaining question to address is to determine the amount of strain to achieve such population, which is the focus of the following section.
  
\section{Valley population of strained donors - Fourier analysis}

We analyze the presence of $x$ and $y$ valleys in the state using Fourier analysis of the tunneling current around the first Brillouin zone, as resonances arise due to the presence of both lattice and valley frequencies in the donor ground state~\cite{Saraiva2016}. In particular, resonances can be expected around $k_x,k_y \sim0.81k_0$ (with $k_0{=}2\pi/a_0$), which are valley interference resonances only, with an amplitude proportional to the $x$-and $y$-valley population, respectively. The 2D Fourier transform of the experimental images of the $D^0$ and $D^-$ state centred on the first Brillouin zone are shown in Fig.~\ref{fig4}a and b, respectively. A valley signal can be observed for both images around these frequencies, as indicated by the green ellipsoids. We show in Fig.~\ref{fig4}c the Fourier transform of the theoretical $D^0$ image at 20\%-Ge strain. In agreement with the experimental Fourier transforms, the theoretical one shows a strong signal at the lattice frequencies, but importantly it also contains a remaining signal at the valley frequencies. The presence of a residual $xy$-valley population observed experimentally for such large strain is not expected from effective mass (EM) theory~\cite{Wilson1961,Koiller2002} as explained below. Following this EM framework, an energy shift of the valley states is associated with a change of the lattice constants upon strain, which can be characterised by a valley strain parameter $\chi$ linked to the deformation potential $\Xi_u{\sim}8.6$\,eV and the valley-orbit parameter $\Delta_c{\sim}2$\,meV for silicon~\cite{Wilson1961, Koiller2002}:

\begin{equation}
\begin{gathered}
\chi=\frac{\Xi_u}{3\Delta_c}\frac{a_{Si}-a_{Ge}}{a_{Si}}(\frac{2c_{12}}{c_{11}}+1)x
\label{chi}
\end{gathered}
\end{equation}

\noindent This energy shift of the valley states causes the donor ground state to evolve from the the so-called $A_1$ state, where all the valleys are equally distributed, to the $A_{1z}^\infty$ state made out $z$-valleys only (see Supporting Information section S5). The evolution of the $z$-valley population according to the EM model is plotted in Fig.~\ref{fig4}d as a function of the Ge concentration (see Supplementary Information), and it can be compared to the valley population obtained from our TB atomistic simulations. The $z$-valleys quickly dominate (above 90\%) for strain larger than 3.2\% in the EM case, to exceed 99.7\% at 20\%-Ge, while the valley repopulation is much slower in the the TB case with a remaining 4.6\% of $xy$-population at 20\%-Ge, which qualitatively match our experimental observation.\\

As depicted in Fig.~\ref{fig1}a, the evolution of the valley population is directly related to the energy difference between the valley states, which is a good starting point to gain a better understanding of why the change in valley population as calculated by TB is slower than by EM. Therefore, we show in Fig.~\ref{fig4}e the evolution of the state energies as a function of strain for both EM and TB calculations. Both models show a qualitatively similar behaviour for the $1s$-manifold (black lines for EM, blue dots for TB), with a large first excited state energy (about 12\,meV) at zero strain which decreases to a few meV at finite strain. Both models also show an increasing splitting between the $z$-valley and $xy$-valley states (black and blue arrows for the EM and TB model, respectively) as the Ge content (hence the strain) is increased. However, the energy splitting between these valley states is smaller for TB than it is for EM calculations for the same amount of strain (see arrows), which drives a slower valley repopulation of the ground state. In order to understand this quantitative difference between the two models, it is important to note that the EM model only considers the first $1s$-like manifold, while TB takes into account a larger donor spectrum basis including higher orbital states~\cite{Lansbergen2011,Tankasala2018} (e.g., $2s$, $2p$ and so on). In Fig.~\ref{fig4}e, some of these higher orbital states (red dots) can be seen coming down in energy with strain and crossing the $xy$-valley states, which can reduce the energy gap between the $z$ and $xy$-valley states in turn driving the valley population. As future work, it would be relevant to extend the EM model to include higher orbital states in the calculations to see if it converges towards the TB ones, as well as to further investigate in the TB framework the interaction and possible hybridization between the $1s$-like and higher orbital states when they cross around 5\% Ge concentration.\\

\section{Conclusion}

In conclusion, we have engineered a thin epilayer of highly strained silicon to embed and image the wave functions of single phosphorus donors. Our low-temperature fabrication procedure allows to maintain a high level of strain, as confirmed by X-ray diffraction, and to obtain a high quality surface for STM purposes. We evidence a clear dominance of the $z$-valley population in the STM image in contrast to the relaxed case, which can be used to make the exchange and tunnel coupling isotropic in the $xy$-plane. A Fourier analysis highlights the residual presence of $xy$-valley states for large strain that well exceeds predictions based on EM theory. This effect can be explained using atomistic TB simulations showing the influence of higher orbital states, not previously considered in EM calculations. These results provide the necessary understanding of strain mechanisms and impact on donor quantum states in silicon necessary to harness their potential in view of quantum nanoelectronics.

\section{Acknowledgments}
B.V., K.N., J.S., M.U, A.T., B.C.J., J.C.MC., M.Y.S., R.R., L.C.L.H. and S.R. acknowledge support from the ARC Centre of Excellence for Quantum Computation and Communication Technology (CE170100012). J.S. acknowledges support from an ARC DECRA fellowship (DE160101490). This work was in part funded by the U.S. Army Research Office (W911NF-17-1-0202). B.C.J. and J.C.MC. acknowledge the AFAiiR node of the NCRIS Heavy Ion Capability for access to ion-implantation facilities. C.W. and N.V. acknowledge support from ARC Centre of Excellence Future Low Energy Electronics Technologies Grant CE170100039. Computational resources were provided by the National Computing Infrastructure (NCI) and Pawsey Supercomputing Center through National Computational Merit Allocation Scheme (NCMAS) and from the LIEF HPC-GPGPU facility hosted at the University of Melbourne with the assistance of LIEF Grant LE170100200.
\bibliography{strain.bib}

\setcounter{equation}{0}
\setcounter{figure}{0}
\setcounter{table}{0}
\renewcommand{\theequation}{S\arabic{equation}}
\renewcommand{\theHequation}{S\arabic{equation}}

\renewcommand{\thefigure}{S\arabic{figure}}
\renewcommand{\theHfigure}{S\arabic{figure}}

\clearpage
\onecolumngrid

\begin{widetext}
\begin{center}
\large{\textbf{Valley population of spatially resolved donor states in highly strained silicon\\-\\Supporting Information}}
\end{center}
\end{widetext}

\onecolumngrid

\section{S1 - X-ray diffraction analysis}

The X-ray diffraction measurements were performed along the (224) crystallographic axis with a wavelength $\lambda{=}0.15418$\,nm. The measured $2\theta-\Omega$ maps are converted to real-space maps by using the following relationships:

\begin{equation}
\begin{gathered}
a_{\parallel}{=}2\sqrt{2}\lambda\big[\cos(\Omega')-\cos(2\theta'-\Omega')\big]^{-1}\\
a_{\perp}{=}4\lambda\big[\sin(\Omega')+\sin(2\theta'-\Omega')\big]^{-1}\\
\theta'{=}\theta{-}\theta_0 \quad , \quad \Omega'{=}\Omega{-}\Omega_0
\label{xray:coo}
\end{gathered}
\end{equation} 

\noindent where $\theta_0$ and $\Omega_0$ are two offset constants interpolated from the maximum of each map, corresponding to the response of the unstrained substrate of the sSOI wafer with the silicon lattice constant $a_{\parallel}{=}a_{\perp}{=}0.5431$\,nm.\\

The X-ray diffraction signal corresponding to the strained epilayer is fitted in real-space to a 2D Gaussian:

\begin{equation}
\begin{gathered}
A(a_{\parallel},a_{\perp}){=}A_{off}+A_0e^{-\big(\frac{a_{\parallel}-a_{\parallel0}}{\sqrt{2}c_{\parallel}}\big)^2-\big(\frac{a_{\perp}-a_{\perp0}}{\sqrt{2}c_{\perp}}\big)^2}
\label{xray:fit}
\end{gathered}
\end{equation}

\begin{figure*}[htp]
\begin{center}
\includegraphics[width=\textwidth]{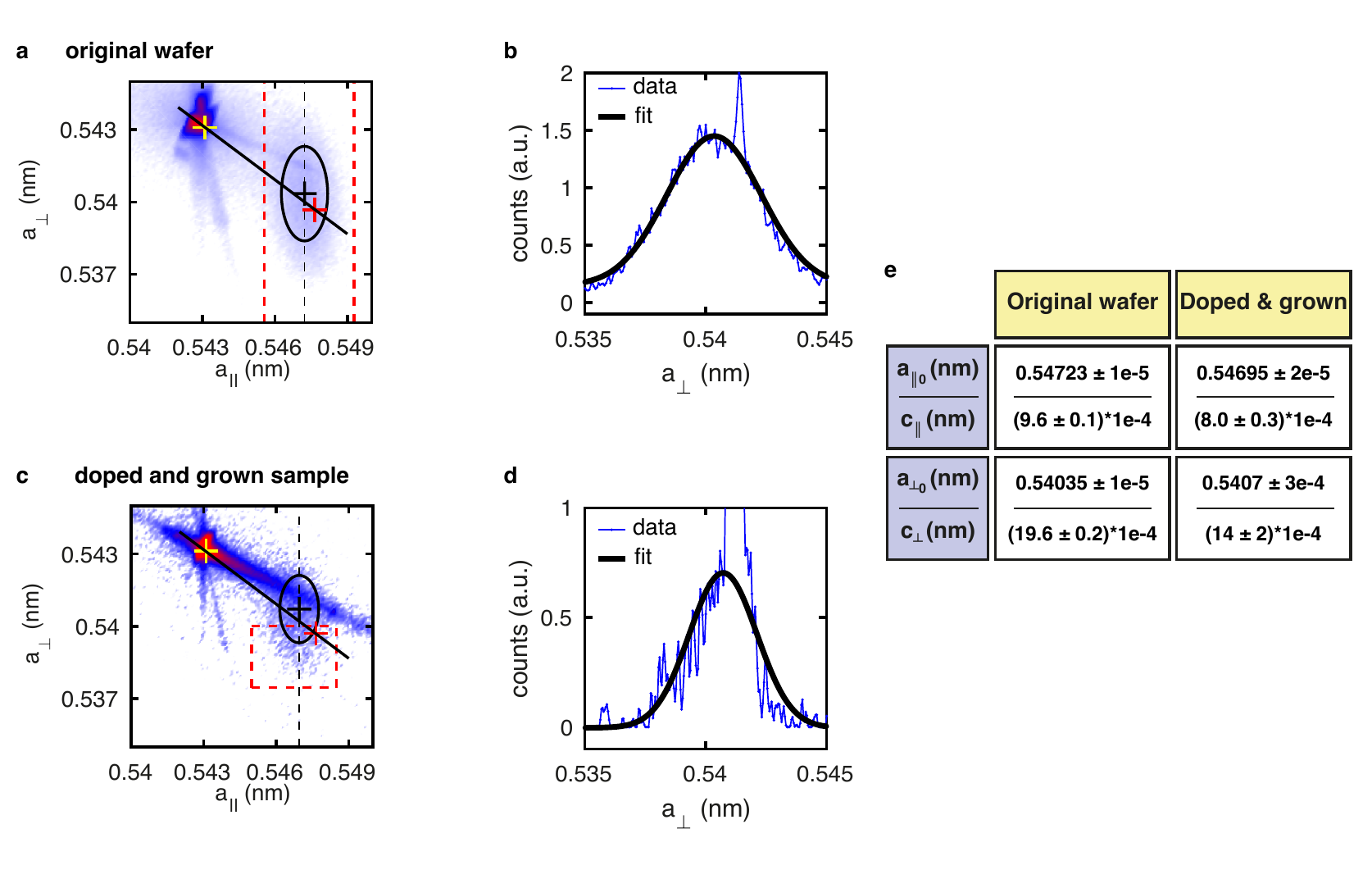}
\caption{\textbf{X-ray diffraction maps - Fitting parameters. a,} Real space map of the original strained SOI wafer, same as main text. The red dashed lines indicate the boundaries chosen for fitting the signal corresponding to the strained epilayer. \textbf{b,} Cut along the black dashed line of the map shown in \textbf{a} (blue), with the resulting Gaussian fit (black line).  \textbf{c-d} Same as \textbf{a-b} for the doped and overgrown sample. Because of the weaker strain signal and the strong artifact line, only the bottom part of the strain signal was considered for the fit. \textbf{e,} Table summarising the values obtained from the fits.}
\label{figS2}
\end{center}
\end{figure*}

Details about the fits are given in Fig.~\ref{figS2}. Only the lower part of the signal was considered for the fit of the doped and overgrown sample in order to avoid the artifact line due to the X-ray measurement setup. Line cuts and corresponding fits are shown in Fig.~\ref{figS2}b and d, for the original wafer and measured sample, respectively. A table summarising the values obtained from the fits is shown in Fig.~\ref{figS2}e.

\section{S2 - Thermal broadening fit of the tunnel current}

The tunnel current $I_{sd}(V_b)$ measured over the donor, as shown in Fig.\ref{fig3}a, is fitted to the following double resonance equation in the thermal broadening regime:

\begin{equation}
I(V_b)=\sum_{i=D^0,D^-} [I_0^i+\Delta_{\Gamma}^i(V_{b0}^i-V_b)]\int_0^{|V_b|}{cosh^{-2}\left(\frac{\alpha^i e (U-|V_{b0}^i|)}{2k_BT}\right)dU}
\label{eq:tunnel current}
\end{equation}

\noindent where $k_B$ is the Boltzmann constant, T the temperature fixed to 4.2\,K, $\alpha_i$ the lever-arm parameters, $V_{b0}^i$ the resonance thresholds, $I_0^i$ the amplitudes, and $Delta_{\Gamma}^i$ a vacuum barrier lowering parameter, with $i$ corresponding to either the $D^0$ or the $D^-$ resonance. The parameters and error bars obtained from the fit are shown in Table~\ref{tab:TunnelFit}.\\

\begin{table}[h]
\caption{Parameters obtained by fitting the bias spectroscopy data to Eq.~\ref{eq:tunnel current} corresponding to a thermal broadening regime.} 
\centering 
\begin{tabular}{c c c} 
\hline\hline 
 Resonance & Parameter & Fit value
\\ [0.5ex]
\hline 
\\ [0.5ex]
& $V_{b0}^0$\,(V) & -0.380$\pm$0.001 \\
& $\alpha^0$\,(\%) & 6.5$\pm$0.9 \\
\raisebox{1.5ex}{$D^0$}& $I_0^0$\,(pA) & 7.4 $\pm$0.1\\
& $\Delta_\Gamma^0$\,(pA/V) & 0.9$\pm$0.6 \\[5ex]
& $V_{b0}^-$\,(V) & -0.779$\pm$0.002 \\
& $\alpha^-$\,(\%) & 3.4$\pm$0.4 \\
\raisebox{1.5ex}{$D^-$}& $I_0^-$\,(pA) & 7.2$\pm$0.3 \\
& $\Delta_\Gamma^-$\,(pA/V) & 17.2$\pm$1.6 \\[5ex]
\hline 
\end{tabular}
\label{tab:TunnelFit}
\end{table}

The charging energy $E_C$ between the first and the second electron is defined as:
\begin{equation}
E_C{=}\alpha^-(|V_{b0}^-|-|V_{b0}^0|)
\label{eq:charging}
\end{equation}

\section{S3 - Theoretical STM images vs donor depth}

\begin{figure*}[htp]
\begin{center}
\includegraphics[width=\textwidth]{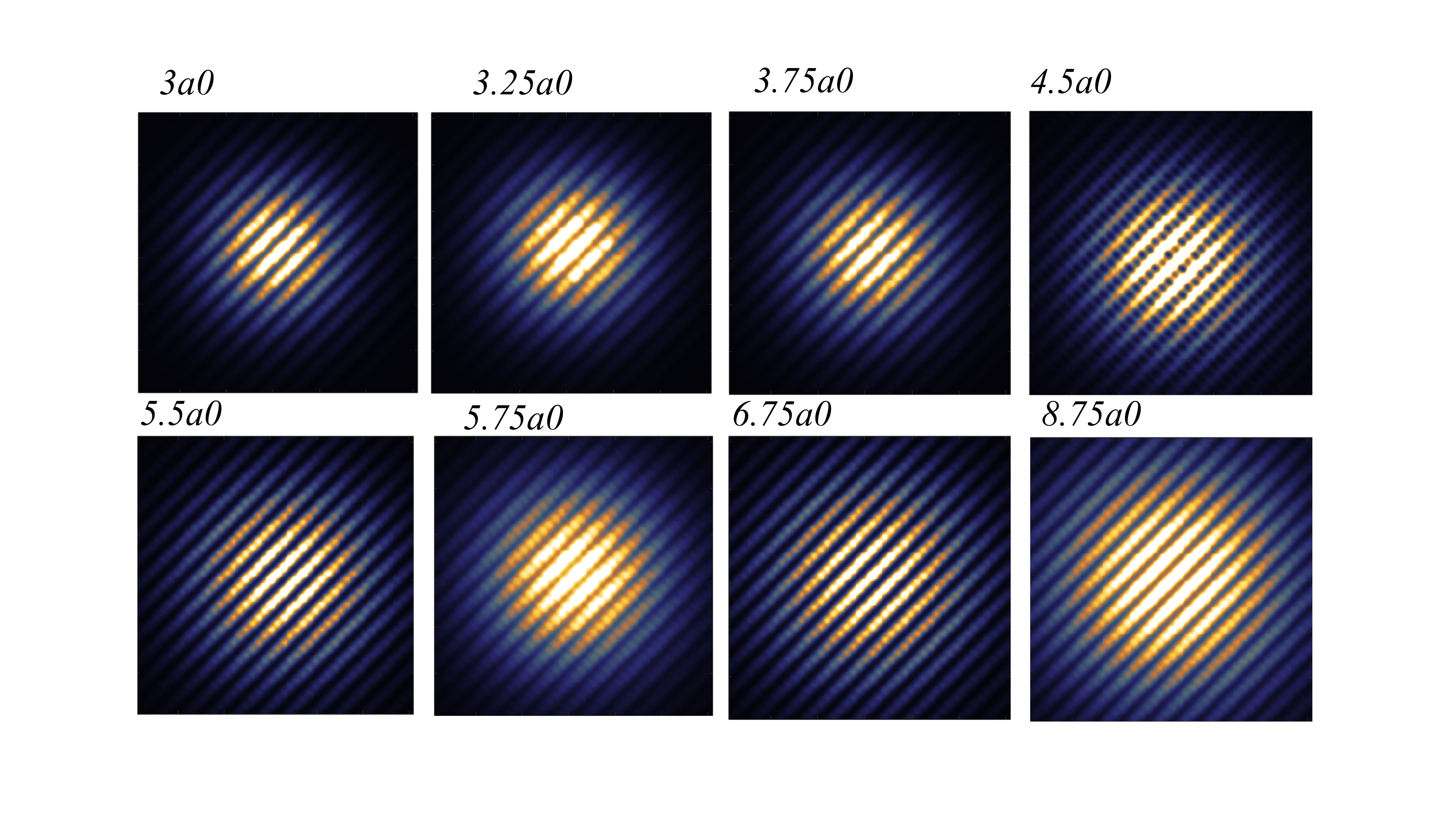}
\caption{\textbf{Theoretical STM images - Depth dependence.} The images are calculated for 20\%-equivalent Ge strain according to the same tight-binding framework as described in the main text. The images look qualitatively the same due to the absence of $xy$-valley components in the ground state, so the images are independent of the donor position with respect to the surface dimers. The difference in contrast comes from the $z$-valley interference condition at the silicon surface.}
\label{figS4}
\end{center}
\end{figure*}

We have used the same theoretical framework as described in the main text to compute theoretical images STM of strained donors (20\% Ge) for depths ranging from $3\,a_0$ to $8.75\,a_0$. The results are shown in Fig.~\ref{figS4}. The characteristic oscillating pattern observed in the unstrained case disappears for all the depths.

\section{S4 - Heitler-London exchange and tunnel coupling in the effective mass model}

This section describes the effective mass model used for the tunnel coupling calculations shown in Fig.~\ref{fig5}. We start from the expression of the exchange interaction between two donors in the Heitler-London (HL) regime~\cite{Koiller2001,Voisin2020}, assuming an effective mass expression of the single donor ground state:

\begin{equation}
\begin{gathered}
\psi(\vec{r})=\sum_{\mu =1}^6 \alpha_{\mu}F_{\mu}(\vec{r})\phi_{\mu}(\vec{r}) \\
with \quad F_{\pm z}(\vec{r})=\frac{1}{\sqrt{\pi a^2b}}e^{-\sqrt{\frac{(x^2+y^2)}{a^2}+\frac{z^2}{b^2}}} \quad \textrm{anisotropic \; envelope}\\
and \quad \phi_{\mu}(\vec{r})=u_{\mu}(\vec{r})e^{i\vec{k_{\mu}}\cdot \vec{r}} \quad \textrm{Bloch \; functions}
\end{gathered}
\end{equation}

The $\mathrm{\alpha_{\mu}}$ represent the valley population distribution among the 6 valleys $\pm x$, $\pm y$ and $\pm z$. $a$ and $b$ represent the transverse and longitudinal Bohr radii, respectively, and $k_{\mu}$ is the valley momentum. For large enough distance between the phosphorus atoms, the singlet-triplet splitting is dominated by the exchange integral, which can be defined in the HL approximation as:\\

\begin{equation}
\begin{gathered}
J(\vec{R})=\iint \vec{dr_1}\vec{dr_2}\psi^*(\vec{r_1})\psi^*(\vec{r_2}-\vec{R})\frac{e^2}{\epsilon |\vec{r_1}-\vec{r_2}|}\psi(\vec{r_1}-\vec{R})\psi(\vec{r_2})
\end{gathered}
\end{equation}

This expression can be simplified using a couple of approximations, namely neglecting the inter-valley exchange terms~\cite{Salfi2018} and assuming that the two donors have the same valley population distribution. We obtain:

\begin{equation}
\begin{gathered}
J(\vec{R})= A \sum_{\mu \nu} \alpha_{\mu}^2\alpha_{\nu}^2 \; j_{\mu \nu}(\vec{R})\; \cos((\vec{k_{\mu}}-\vec{k_{\nu}})\cdot \vec{R})\\
with \quad j_{\mu \nu}(\vec{R})=\iint \vec{dr_1}\vec{dr_2}\tilde{j}_{\mu \nu}(\vec{R})=\iint \vec{dr_1}\vec{dr_2}F_{\mu}^*(\vec{r_1})F_{\nu}^*(\vec{r_2}-\vec{R})\frac{e^2}{\epsilon |\vec{r_1}-\vec{r_2}|}F_{\mu}(\vec{r_1}-\vec{R})F_{\nu}(\vec{r_2})
\label{exc:EM}
\end{gathered}
\end{equation} 

The envelope terms can be simplified assuming $F_{\mu}(\vec{r_1})F_{\nu}(\vec{r_2}) \sim F_{\mu}(\vec{r_1})F_{\nu}(\vec{r_2})\delta(\vec{r_1}-\vec{r_2})$ because of the exponential nature of the orbitals, which finally leads to:

\begin{equation}
\begin{gathered}
j_{\mu \nu}(\vec{R}) = \frac{1}{|\vec{R}|}F_{\mu}(\vec{R})F_{\nu}(\vec{R})F_{\mu}(\vec{0})F_{\nu}(\vec{0})\\
\end{gathered}
\end{equation}

\begin{figure*}[htp]
\begin{center}
\includegraphics[width=\textwidth]{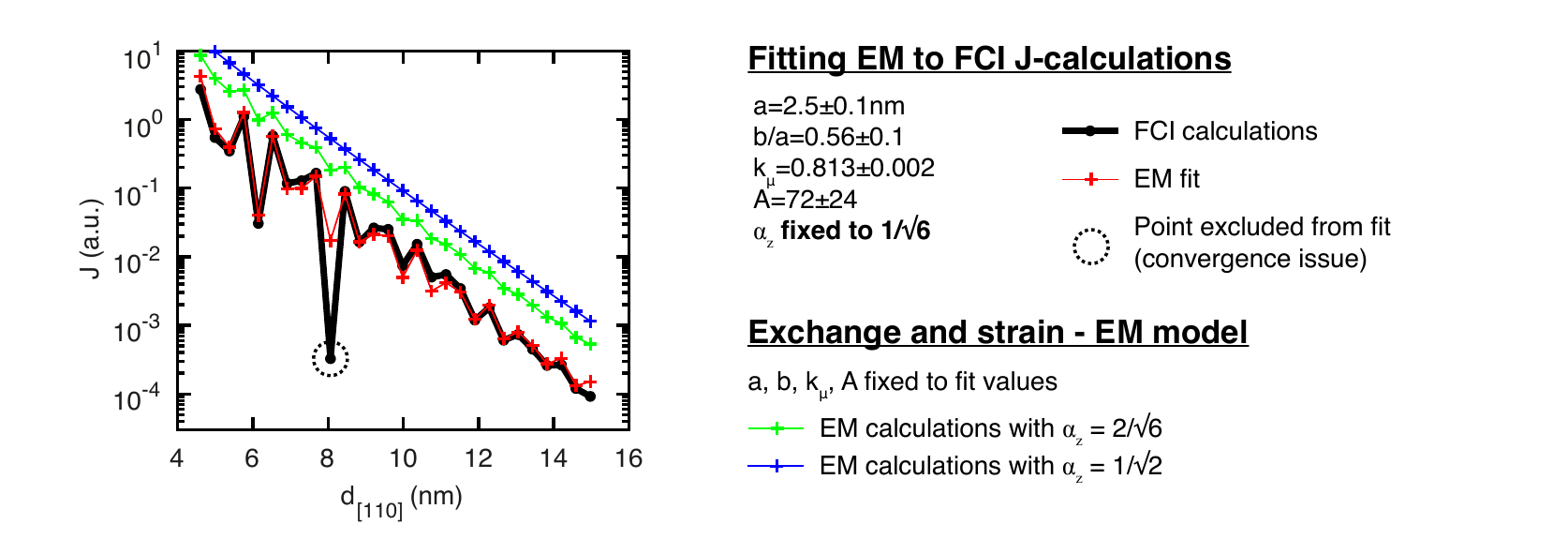}
\caption{\textbf{Exchange and $z$-valley population. FCI and EM models.} Exchange calculations obtained via an atomistic FCI method for bulk donors (black dots), and EM model fit with a valley population fixed to 33\% (red crosses). The circled data point was excluded from the fit. The fitted parameters $a$, $b$, $k_\mu$ and $A$ are then fixed and the $z$-valley population is increased to 66\% and 100\% (green and blue crosses, respectively). As the $z$-valley population increases, the exchange variations due to in-plane valley interference vanish and the exchange increases since the $j_{zz}$ terms, which are fully constructive in the plane, dominate  the exchange expression along [110].}
\label{figS3}
\end{center}
\end{figure*}

To test the validity and calibrate the parameters $a$, $b$, $k_{\mu}$ and $A$ of this expression, we have fixed the valley population to 33\% (i.e. $\alpha_z{=}1/\sqrt{6}$) for each valley and fitted full configuration interaction (FCI) exchange calculations along [110] obtained by solving directly the full $2e$-Hamiltonian, including Coulomb and electron interactions, using a 20-orbital spin resolved basis set per atom obtained from atomistic simulations of the $1e$-problem for bulk donors in unstrained silicon. The FCI exchange and the calibrated EM model are shown in Fig.~\ref{figS3}, which shows a good agreement down to inter-donor distances of 5\,nm. The values $b/a$ and $k_\mu$ obtained from the fit are in agreement with experimental data found in ref.~\cite{Voisin2020}.\\

To model the influence of the valley population on exchange and tunnel couplings, we fix the parameters $a$, $b$, $k_{\mu}$ and $A$, and only vary the valley population weights $\alpha_\mu$, while respecting the plane symmetry $\alpha_x{=}\alpha_y$ and the normalisation $\sum{\alpha_\mu^2}{=}1$. We show in Fig.~\ref{figS3} the exchange values obtained for 66\% and 100\% $z$-valley population (green and blue crosses, respectively). The large variations predicted for the bulk case, which are due to in-plane valley interference of the $x$ and $y$-valleys, are washed out at larger $z$-valley population and the overall exchange amplitude also increases. Both effects relate to the $j_{zz}$ terms becoming more and more dominant along [110] out of the 36 terms of Eq.~\ref{exc:EM} as strain increases, and these terms are fully constructive in the $xy$-plane~\cite{Voisin2020}. For each configuration, the tunnel coupling is derived from the exchange energy through the Heitler-London equation~\ref{exc} of the main text, assuming a constant charging energy constant with the inter-donor distance.

\section{S5 - Valley repopulation in the effective mass model}

We derive the valley population of the first six $1s$-like donor states following the effective mass model developed in Ref.~\onlinecite{Wilson1961} and used in Ref.~\onlinecite{Koiller2002,Usman2015} in the unstrained case. We label $\Delta_c$ the inter-valley scattering processes energy between valleys of different axis and $\Delta_c(1+\delta)$ for valleys of opposite sign of the same axis. These inter-valley processes lift the six-fold degeneracy of the ground state of an electron bound to a single donor in silicon. In the $\{\pm z,\pm x,\pm y\}$ valley basis, this valley-orbit Hamiltonian reads:

\begin{equation}
\begin{gathered}
H_{vo}=-\Delta_c
\begin{pmatrix}
0 & (1+\delta) & 1 & 1 & 1 & 1\\
(1+\delta) & 0 & 1 & 1 & 1 & 1\\
1 & 1 & 0 & (1+\delta) & 1 & 1 \\
1 & 1& (1+\delta) & 0 & 1 & 1 \\
1 & 1 & 1 & 1 & 0 & (1+\delta) \\
1 & 1 & 1 & 1 & (1+\delta) & 0 \\
\end{pmatrix}
\end{gathered}
\end{equation}

The eigenstates and eigenvalues of $H_{vo}$ are given in Table.~\ref{tab:unstrained}. The ground state is a singlet called $A_1$ with an equal distribution of the wave function across the six valleys. Using $\Delta_c{=}2.16$\,meV and $\delta=-0.3$ reproduces the energy splittings experimentally measured for phosphorus donors. A triplet called $T_2$, with energy $\Delta_c(1+\delta)$, and a doublet called $E$, with energy $\Delta_c(1-\delta)$, follow.

\begin{table}[h]
\caption{Energies and valley population of the first six $1s$-like donor states in the unstrained case.} 
\centering 
\begin{tabular}{c c c l} 
\hline\hline 
 State & & Energy & Valley population
\\ [0.5ex]
\hline 
\\ [0.5ex]
Singlet & $A_1$ & $-\Delta_c(5+\delta)$ & $1/\sqrt{6} (1,1,1,1,1,1)$\\[5ex]
& $T_{2z}$ &  & $1/\sqrt{2}(1,-1,0,0,0,0)$\\
Triplet & $T_{2x}$ & $\Delta_c(1+\delta)$ & $1/\sqrt{2}(0,0,1,-1,0,0)$\\
& $T_{2y}$ &  & $1/\sqrt{2}(0,0,0,0,1,-1)$\\[5ex]
& $E_{z{-}xy}$ &  & $1/\sqrt{12}(2,2,-1,-1,-1,-1)$\\[-1ex]
\raisebox{1.5ex}{Doublet} & $E_{xy}$ & \raisebox{1.5ex}{$\Delta_c(1-\delta)$} & $1/2(0,0,1,1,-1,-1)$\\[2ex]
\hline 
\end{tabular}
\label{tab:unstrained}
\end{table}

Applying strain lifts the valley degeneracy, which can be expressed as a shift of the valley energies via the strain parameter $\chi$, with $\chi{<}0$ for the case of in-plane tensile strain considered here. The perturbed Hamiltonian reads:

\begin{equation}
\begin{gathered}
H_{strain}=H_{vo}+\Delta_c
\begin{pmatrix}
2\chi & 0&0&0&0&0\\
0&2\chi & 0&0&0&0\\
0&0&-\chi &0&0&0\\
0&0&0&-\chi & 0&0\\
0&0&0&0&-\chi &0\\
0&0&0&0&0 &-\chi \\
\end{pmatrix}
\end{gathered}
\end{equation}

Tensile strain favours the $z$-valleys, and it mixes the states which have a finite valley population in both $z$ and $xy$-valleys, i.e. $A_1$ and $E_{z{-}xy}$. The four other states shift linearly in energy, by $-2\chi \Delta_c$ for $T_{2z}$ and by $\chi \Delta_c$ for $T_{2x}$, $T_{2y}$ and $E_{xy}$, and their valley population remains unaffected.\\

The symmetry of the perturbed Hamiltonian respects that of the unstrained states $A_1$ and $E_{z-xy}$'s valley population, thus we should look for valley distributions of the form $(\alpha_z,\alpha_z,\alpha_{in},\alpha_{in},\alpha_{in},\alpha_{in})$ in order to find the mixed states energies $E_{str}$. Together with a normalisation constant, this leads to the following system of equations:

\begin{equation}
\begin{gathered}
2\chi\alpha_z-(1+\delta)\alpha_z-4\alpha_{in}=E_{str}\alpha_z \\
-2\alpha_z-[\chi+(1+\delta)-2]\alpha_{in}=E_{str}\alpha_{in}\\
4\alpha_{in}^2+2\alpha_z^2=1
\end{gathered}
\end{equation}

\noindent Multiplying the first equation by $\alpha_{in}$, the second one by $\alpha_z$, and subtracting them leads to:

\begin{equation}
\begin{gathered}
2\chi\alpha_z-(1+\delta)\alpha_z-4\alpha_{in}=E_{str}\alpha_z \\
4\alpha_{in}^2-2\alpha_z^2=(3\chi+2)\alpha_{in}\alpha_z\\
4\alpha_{in}^2+2\alpha_z^2=1
\end{gathered}
\end{equation}

\noindent Squaring and subtracting the last two equations leads to:

\begin{equation}
\begin{gathered}
2\chi\alpha_z-(1+\delta)\alpha_z-4\alpha_{in}=E_{str}\alpha_z \\
3\alpha_{in}\alpha_z\sqrt{\chi^2+4\chi/3+4}=\pm1\\
4\alpha_{in}^2+2\alpha_z^2=1
\end{gathered}
\end{equation}

\noindent The last two equations can be combined to obtain the expressions for $\alpha_{in}^2$ and $\alpha_z^2$ and the associated energies $E_{str}$ follows, which read:

\begin{equation}
\begin{gathered}
\alpha_z^2=\frac{1} {4} \Big[ 1 \mp \frac{\chi+2/3}{\sqrt{\chi^2+4\chi/3+4}} \Big]\\
\alpha_{in}^2=\frac{1} {8} \Big[ 1 \pm \frac{\chi+2/3}{\sqrt{\chi^2+4\chi/3+4}} \Big]\\
E_{str}=\Delta_c \Big[ \chi/2-(2+\delta)\mp \frac{3}{2}\sqrt{\chi^2+4\chi/3+4} \Big]
\label{eq:valley pop}
\end{gathered}
\end{equation} 

\begin{figure*}[htp]
\begin{center}
\includegraphics[width=\textwidth]{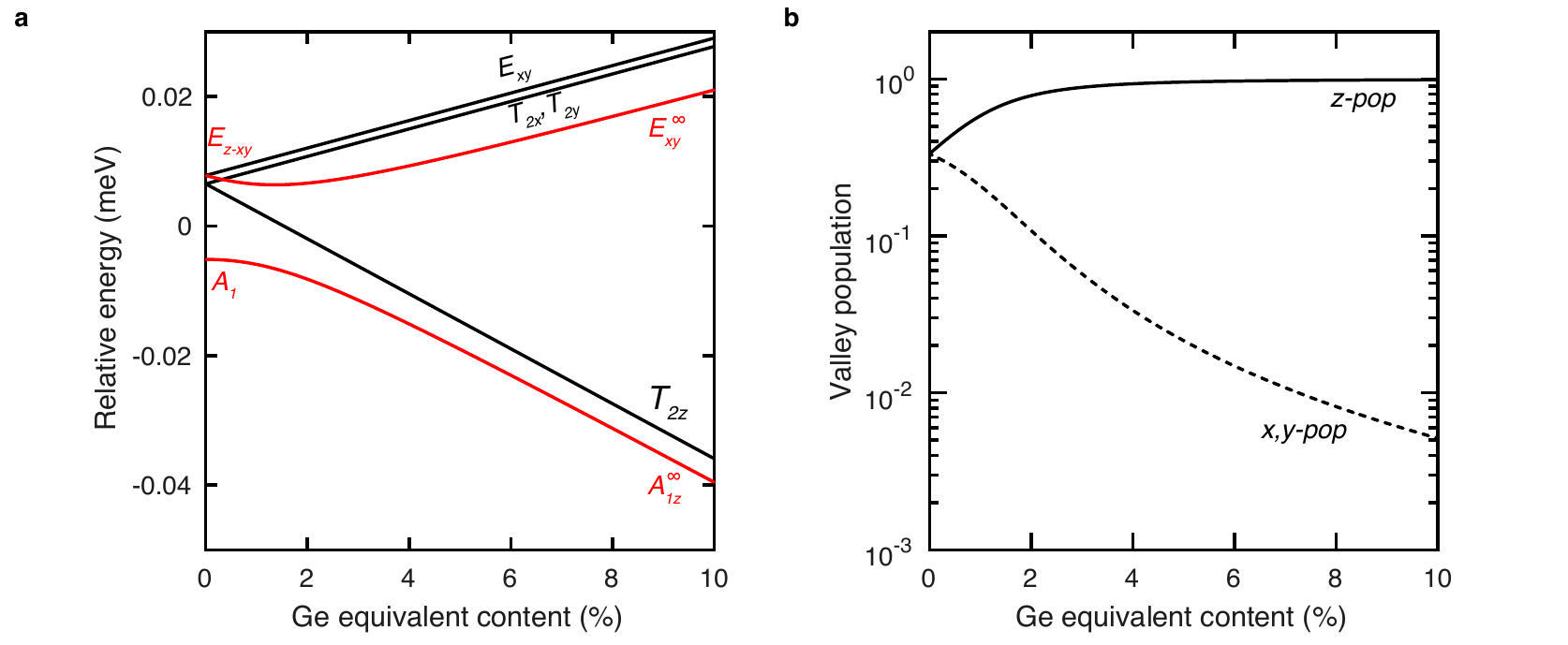}
\caption{\textbf{Effective mass modelling of single donor state energies and valley population vs strain. a,} Evolution of the donor state energies upon strain. The valley degeneracy is broken upon tensile strain as the $z$-valley states are favoured in energy. The $A_1$ state evolves to a $A_{1z}^\infty$ state. The degeneracy between the $T_2$ states is lifted, with $T_{2z}$ which becomes the first excited state about 3\,meV above the $A_{1z}^\infty$ state. The red lines indicate the two states which hybridize upon strain. \textbf{b,} Evolution of the valley population of the donor ground state ($A_1$ to $A_{1z}^\infty$) following Eq.~\ref{eq:valley pop}.}
\label{figS1}
\end{center}
\end{figure*}

\noindent The upper (respectively lower) sign corresponds to the expressions for the ground (respectively excited) state, and its total $z$-valley population (i.e., $2\alpha_z^2$), is the quantity plotted in Fig.\ref{fig4}d. The parameter $\chi$ can be expressed as a function of the Ge content $x$ following Eq.~\ref{chi}, which yields $\chi{\sim}-98.2 x$ using the elastic constants of silicon $\Xi_u{\sim}8.6$\,eV, $\Delta_c{\sim}2.16$\,meV, $c_{11}{=}166$\,GPa and $c_{12}{=}64$\,GPa. The evolution of the valley states energies and valley population is summarised in Fig.~\ref{figS1}. In the asymptotic limit for large Ge content (i.e., $\chi{\rightarrow}{-}\infty$), the ground state is only populated in the $z$-valleys called $A_{1z}^{\infty}=1/\sqrt{2}(1,1,0,0,0,0)$ and $T_{2z}$ has become the first excited state with a splitting equal to $2\Delta_c(1+\delta){\sim}3$\,meV. In the asymptotic limit, $E_{z{-}xy}$ has evolved to be only populated in the $xy$-valleys, called $E_{xy}^\infty{=}1/2(0,0,1,1,1,1)$.

\clearpage

\clearpage

\clearpage

\clearpage

\end{document}